\documentclass[12pt,verbose=true,letterpaper]{article}
\usepackage[margin=1in, top = 1in, bottom = 1in]{geometry}
\usepackage{setspace}
\usepackage{longtable}
\usepackage{makecell}
\usepackage{multirow}
\usepackage{array}
\usepackage[table]{xcolor}
\usepackage[toc,page]{appendix}
\def\keywordname{{\bfseries \emph Keywords}}%
\def\keywords#1{\par\addvspace\smallskipamount{\rightskip=0pt plus1cm
\def\and{\ifhmode\unskip\nobreak\fi\ $\cdot$
}\nopagebreak\noindent\keywordname\enspace\ignorespaces#1\par}}

\usepackage[utf8]{inputenc} 
\usepackage[T1]{fontenc}    
\usepackage{hyperref}       
\usepackage{url}            
\usepackage{booktabs}       
\usepackage{amsfonts}       
\usepackage{nicefrac}       
\usepackage{microtype}
\usepackage{lipsum}
\usepackage{fancyhdr}       
\usepackage{graphicx}       
\usepackage[english]{babel}
\usepackage{authblk}
\usepackage[numbers,sort&compress, square]{natbib}
\usepackage{amsmath}
\usepackage{color}
\usepackage{bm}
\newtheorem{theorem}{Theorem}
\title{\vspace{-2cm}Toxicity Monitoring Rule for a Two-Cohort Phase II Clinical Trial with Bivariate Beta Prior}

\author{
  Yu Wang, Aniko Szabo \\
  Medical College of Wisconsin \\
  Milwaukee, WI 53222\\
  \texttt{aszabo@mcw.edu} \\
}

\begin{document}
\maketitle

\begin{abstract}
Toxicity monitoring is essential in Phase II clinical trials to ensure participant safety. While monitoring rules are well-established for single-arm trials, two-cohort trials present unique challenges because toxicities are expected to be similar between cohorts but may still differ. Current approaches either monitor the two cohorts independently, which ignores their similarity, or pool them together as a single arm, which neglects heterogeneity between cohorts. We propose a Bayesian method based on a bivariate beta prior that provides a compromise between these two approaches. The marginal posterior distribution is derived as a mixture of beta distributions, enabling exact calculations of the proposed method's operating characteristics. Examples demonstrate that joint monitoring offers a balanced approach between the independent and pooled methods.
\end{abstract}

\pagebreak
\section{Introduction}

Phase II trials serve as a critical bridge between Phase I dose-finding studies and Phase III confirmatory trials. In these trials, investigators seek preliminary evidence of treatment efficacy at the dose level recommended from Phase I trials \cite{ivanovaContinuousToxicityMonitoring2005}. However, toxicity monitoring remains essential because the small sample sizes in Phase I trials may result in an inaccurate maximum tolerated dose (MTD), potentially exposing patients to excessive toxicity in subsequent trials. Effective toxicity monitoring involves tracking the number of adverse events relative to the number of enrolled subjects as the trial progresses. When toxicity exceeds a predetermined threshold, the trial must be evaluated immediately and terminated or modified as necessary.

Patient heterogeneity often necessitates dividing the Phase II trial population into two or more cohorts that receive the same intervention. While these cohorts may experience different levels of efficacy, researchers expect them to have similar toxicity profiles. For example, in a breast cancer clinical trial, patients with triple-negative breast cancer may experience different outcomes compared to those with other subtypes. However, when treated with the same therapeutic agent, they are likely to encounter similar side effects. Although the underlying biology of different subtypes varies, patients may still share comparable adverse effects due to the systemic nature of treatment. Adverse effects can manifest in various organs or tissues, leading to similar toxicity levels even when treatment efficacy differs across subtypes.

Two approaches are commonly employed to monitor toxicity in two-cohort studies. The first approach treats the cohorts as independent sub-studies, neglecting the prior expectation of similar toxicity profiles. The second approach combines the cohorts into a single group, despite potential differences in toxicity, and monitors them together. Both approaches have limitations: the independent approach fails to leverage the expected similarity between cohorts, while the pooled approach ignores heterogeneity between them. In this paper, we propose a Bayesian compromise method that integrates both approaches by explicitly modeling the correlation between cohort toxicities.

\section{Background}\label{sec2}
\subsection{Bayesian Monitoring for One Arm}
\label{s:onearm}

Bayesian monitoring for single-arm Phase II trials is well-established \cite{thallPracticalBayesianGuidelines1994}. Although originally developed for treatment efficacy, these guidelines can be adapted for toxicity monitoring with minimal modifications. Let $X_i, i=1,\ldots, n$ denote the binary toxicity outcome for patient $i$, where $X_i=1$ indicates an adverse event and $X_i=0$ indicates no adverse event. Denoting the probability of toxicity as $\theta$, $X_i$ follows a Bernoulli distribution $Bern(\theta)$, while the sum of $X_i$ follows a binomial distribution $Bin(n,\theta)$. The conjugate prior Beta distribution $Be(a,b)$ is typically chosen for the binomial parameter because it yields a closed-form posterior distribution with an intuitive interpretation.

Suppose $k$ toxicity events are observed among $n$ enrolled subjects. The posterior distribution of $\theta$ follows a $Be(k+a, n-k+b)$ distribution with posterior mean $(k+a)/(n+a+b)$. This result has a meaningful interpretation: the prior contributes $a$ imaginary toxicity events and $b$ imaginary no-toxicity events to the posterior distribution, representing an effective sample size (ESS) of $a+b$ subjects.

Based on these distributional results, toxicity monitoring proceeds by defining a null hypothesis that the toxicity probability $\theta$ is acceptably low, for example, $\theta \le 0.2$. The stopping boundary is then defined such that the trial stops for excess toxicity if the posterior probability $P(\theta \ge 0.2 \mid k \text{ out of } n)$ reaches or exceeds $\tau$, a pre-specified exceedance probability \cite{gellerAdvancesClinicalTrial}. For instance, if the prior effective sample size is 3 patients with $\theta \sim Be(0.6,2.4)$ and $\tau = 0.98$, the stopping boundary would be 3 toxicities out of the first 3 patients, 4 out of 4, 4 out of 5, and so on.

\subsection{Bivariate Beta Distribution}

Olkin et al.\cite{olkinConstructionsBivariateBeta2014} introduced a bivariate beta distribution constructed from the Dirichlet distribution with correlations over the full range $[-1,1]$. Let the variates $U_{11}, U_{10}, U_{01},U_{00}=1-U_{11}-U_{10}-U_{01}$ have a Dirichlet distribution with density
\begin{equation}
f(u_{11},u_{10},u_{01},u_{00}) = \frac{1}{B(\bm \alpha)}u_{11}^{\alpha_{11}-1}u_{10}^{\alpha_{10}-1}u_{01}^{\alpha_{01}-1}(1-u_{11}-u_{10}-u_{01})^{\alpha_{00}-1}
\label{eq:pdfbb}
\end{equation}
for $0\le u_{ij}\le 1, \:i, j=0,1, \: u_{11}+u_{10}+u_{01} \le 1$, $\bm \alpha = (\alpha_{11}, \alpha_{10}, \alpha_{01},\alpha_{00})$, $\alpha_{ij} \ge 0$, and $B(\bm \alpha) = \frac{\Gamma(\alpha_{11})\Gamma(\alpha_{10})\Gamma(\alpha_{01})\Gamma(\alpha_{00})}{\Gamma(\alpha_{11}+\alpha_{10}+\alpha_{01}+\alpha_{00})}$.

The sums
\begin{equation}\label{eq:B2def}
X= U_{11} + U_{10}, \quad Y= U_{11} + U_{01}
\end{equation}
follow a bivariate beta distribution, denoted $[X,Y] \sim \bm {Be}_2(\bm \alpha)$. By the collapsibility property of the Dirichlet distribution, the marginal distributions of $X$ and $Y$ are $Be(\alpha_{1+}, \alpha_{0+})$ and $Be(\alpha_{+1}, \alpha_{+0})$, respectively, where $\alpha_{1+}= \alpha_{11}+\alpha_{10}$, $\alpha_{+1}= \alpha_{11}+\alpha_{01}$, $\alpha_{0+}= \alpha_{01}+\alpha_{00}$, and $\alpha_{+0}= \alpha_{10}+\alpha_{00}$.

The joint density function is
\begin{equation}
f(x,y) = \frac{1}{B(\bm \alpha)}\int_\Omega u^{\alpha_{11}-1}(x-u)^{\alpha_{10}-1}(y-u)^{\alpha_{01}-1}(1-x-y+u)^{\alpha_{00}-1}du
\label{eq:bbeta}
\end{equation}
where $\Omega = \{u: \max(0,x+y-1) < u_{11} < \min(x,y)\}$. Olkin et al.\cite{olkinConstructionsBivariateBeta2014} also derived an explicit expression for the correlation between $X$ and $Y$:
\begin{equation}
\rho(X, Y) = \frac{\alpha_{11}\alpha_{00}-\alpha_{10}\alpha_{01}}{\sqrt{\alpha_{1+}\alpha_{+1}\alpha_{0+}\alpha_{+0}}}.
\label{eq:rho}
\end{equation}

The bivariate beta distribution has been commonly used as a prior distribution for event probabilities in paired binary data. In this work, however, we apply it to unpaired data from two independent cohorts.

\section{Bivariate Beta-Binomial Model}

\subsection{Model Components}
\label{s:model_comp}

Consider independent binomial toxicity counts $X \sim Bin(n_1,\theta_1)$ and $Y \sim Bin(n_2,\theta_2)$ in two cohorts. We propose using the prior $\bm {Be}_2(\bm \alpha)$ for the toxicity probabilities $\theta_1$ and $\theta_2$. Since the two cohorts are independent, the likelihood for observing $X=k_1$ and $Y=k_2$ is
\begin{equation}
L(\theta_1, \theta_2) = {n_1 \choose k_1} \theta_1^{k_1}(1-\theta_1)^{n_1-k_1}{n_2 \choose k_2} \theta_2^{k_2}(1-\theta_2)^{n_2-k_2}.
\label{eq:lik}
\end{equation}
Thus, the posterior density of $[\theta_1,\theta_2]$ is
\begin{equation}
\begin{split}
p(\theta_1, \theta_2|n_1,n_2,k_1,k_2) \propto & \int_\Omega u^{\alpha_{11}-1}(\theta_1-u)^{\alpha_{10}-1}(\theta_2-u)^{\alpha_{01}-1}(1-\theta_1-\theta_2+u)^{\alpha_{00}-1}du\\
& \times\theta_1^{k_1}(1-\theta_1)^{n_1-k_1}\theta_2^{k_2}(1-\theta_2)^{n_2-k_2}.
\end{split}
\label{eq:post}
\end{equation}

\subsection{Marginal Posterior}

The posterior distribution (\ref{eq:post}) does not have a closed form. However, we derive properties that simplify the calculation of posterior probabilities needed for toxicity monitoring and enable a useful interpretation.
\begin{theorem}
\label{thmm:thm1}
The posterior distribution is a mixture of bivariate beta distributions:
\begin{equation}
[\theta_1,\theta_2|n_1,n_2,k_1,k_2]
\sim \sum_{x_1=0}^{k_1}\sum_{x_2=0}^{n_1-k_1}\sum_{y_1=0}^{k_2}\sum_{y_2=0}^{n_2-k_2} w(x_1,x_2,y_1,y_2) \bm {Be}_2\big(\bm \alpha + \bm z(x_1,x_2,y_1,y_2)\big),
\label{eq:joint}
\end{equation}
where $\bm z(x_1,x_2,y_1,y_2) = (x_1+y_1, k_1-x_1+y_2, x_2+k_2-y_1, n_1-k_1-x_2+n_2-k_2-y_2)$, $w(x_1, x_2, y_1, y_2) = c_0 \bm B (\alpha+\bm z) {k_1 \choose x_1} {{n_1-k_1} \choose x_2} {k_2 \choose y_1} {{n_2-k_2} \choose y_2}$, and $c_0$ is a scaling constant.
\end{theorem}

The proof of this theorem is provided in the Appendix.

Analogously to the imaginary event interpretation of the prior for the univariate beta-binomial distribution described in Section \ref{s:onearm}, we can interpret the parameters of this bivariate beta mixture using a potential outcomes framework. For each enrolled subject, we conceptualize two potential outcomes: one if enrolled in cohort 1 and one if enrolled in cohort 2. In practice, only one of these will be observed. Considering both actual and counterfactual outcomes, each subject belongs to one of four groups: those who would experience toxicity in (a) either cohort, (b) only cohort 1, (c) only cohort 2, or (d) neither cohort. Table \ref{tab:fact1} presents the breakdown of observed toxicity counts into these four subcategories.

\begin{table}[htbp]
  \centering
  \caption{Partitions of observed toxicity outcomes based on potential outcomes in two cohorts. The gray background highlights cells corresponding to observed toxicity.}
    \begin{tabular}{lllll}
    \toprule
    {\multirow{2}{4cm}{Potential outcome\\ in Cohort 2}} & {\multirow{2}{3cm}{Source}} & \multicolumn{2}{c}{Potential outcome in Cohort 1} & {\multirow{2}{3cm}{Subtotal}} \\
    \cmidrule{3-4}
    & & Toxicity & Non-Toxicity & \\
    \midrule
    \multirow{3}{2cm}{Toxicity} & Prior & $\alpha_{11}$ & $\alpha_{01}$ & $\alpha_{+1}$ \\
    & Cohort 1 & \cellcolor{lightgray}{$x_1$} & $x_2$ & $x_1+x_2$ \\
    & Cohort 2 & \cellcolor{lightgray}{$y_1$} & \cellcolor{lightgray}{$k_2-y_1$} & \cellcolor{lightgray}{$k_2$} \\
    \midrule
    \multirow{3}{2cm}{Non-Toxicity} & Prior & $\alpha_{10}$ & $\alpha_{00}$ & $\alpha_{+0}$ \\
    & Cohort 1 & \cellcolor{lightgray}{$k_1-x_1$} & $n_1-k_1-x_2$ & $n_1-x_1-x_2$ \\
    & Cohort 2 & $y_2$ & $n_2-k_2-y_2$ & $n_2-k_2$ \\
    \midrule
    \multirow{3}{2cm}{Subtotal} & Prior & $\alpha_{1+}$ & $\alpha_{0+}$ & $ESS$ \\
    & Cohort 1 & \cellcolor{lightgray}{$k_1$} & $n_1-k_1$ & $n_1$ \\
    & Cohort 2 & $y_1+y_2$ & $n_2-y_1-y_2$ & $n_2$ \\
    \bottomrule
    \end{tabular}
  \label{tab:fact1}
\end{table}

For subjects in cohort 1 (rows 2 and 5), we observe only the cohort-1 potential outcome: $k_1$ toxicities among $n_1$ subjects. However, among these $k_1$ patients with toxicity, we posit that $x_1$ would develop toxicity in either cohort, while $k_1-x_1$ would develop toxicity only in cohort 1. Similarly, among the $n_1-k_1$ subjects without toxicity, $x_2$ would develop toxicity only in cohort 2, and $n_1-k_1-x_2$ would experience no toxicity regardless of cohort.

Subjects in cohort 2 (rows 3 and 6) are similarly partitioned, with $y_1$ counting those who would develop toxicity in either cohort among the $k_2$ who actually do, and $y_2$ counting those who would develop toxicity only in the other cohort.

With $\bm\alpha$ representing the number of imaginary subjects encoded in the prior, each component in the mixture (\ref{eq:joint}) represents a specific set of values for $x_1, x_2, y_1, y_2$, with contributions from each source to the four potential-outcome groups. The weights $w(x_1, x_2, y_1, y_2)$ quantify the probability of each combination.

Toxicity monitoring requires marginal posterior distributions, which have a simpler form:

\begin{theorem}
\label{thmm:thm2}
The marginal posterior distribution of $\theta_i$ is a mixture of beta distributions:
\begin{equation}
\begin{split}
[\theta_1|n_1,n_2,k_1,k_2] \sim & \sum_{y=0}^{n_2} g_1(y) \bm{Be}(\alpha_{1+}+k_1+n_2-y,\: \alpha_{0+}+n_1-k_1+y), \\
[\theta_2|n_1,n_2,k_1,k_2] \sim & \sum_{x=0}^{n_1} g_2(x) \bm{Be}(\alpha_{+1}+k_2+n_1-x,\: \alpha_{+0}+n_2-k_2+x).
\end{split}
\end{equation}
The exact forms of $g_1$ and $g_2$ are given in the Appendix.
\end{theorem}

A detailed proof of this theorem based on Theorem \ref{thmm:thm1} is provided in the Appendix. The marginal subtotals in Table \ref{tab:fact1} provide an intuitive justification for this form, where $x=x_1+x_2$ and $y=y_1+y_2$ count the number of subjects in each cohort who would have experienced toxicity in the other cohort.

\section{Two-Cohort Toxicity Monitoring}

The toxicity monitoring framework for two cohorts follows the same principles as single-arm monitoring, but with modifications to account for the correlation between cohorts. We define the null and alternative hypotheses, specify the prior distribution, and set an exceedance threshold for the posterior probability.

\subsection{Hypotheses}

Although we monitor the two cohorts simultaneously, we allow different thresholds $\theta_{0i}$ for acceptable toxicity probability in each cohort. This flexibility is useful when patients in one cohort have a worse prognosis and a higher toxicity level might be acceptable if the treatment provides overall benefit. Thus, we define the hypotheses for each cohort separately:

\[
\text{Cohort } i: \quad H_0: \theta_i \le \theta_{0i} \quad \text{versus} \quad H_1: \theta_i > \theta_{0i} \quad i=1,2.
\]

\subsection{Prior Distribution Parameters}

The bivariate beta-binomial prior described in Section \ref{s:model_comp} has four parameters $\alpha_{ij}, i=0,1, j=0,1$. Rather than specifying these parameters directly, we propose computing them from more interpretable characteristics: the marginal means, the effective sample size $ESS$, and the correlation $\rho$. Based on the expectation that cohorts have similar toxicity, we assume prior means $p_1$ and $p_2$ for the two cohorts, though this is not a necessary restriction. We then solve for $\bm \alpha$ based on $ESS$, $\rho$, and $p$ using the following system of equations:
\begin{equation}
\left\{
\begin{array}{lr}
p_1 = \dfrac{\alpha_{11}+\alpha_{10}}{ESS}, \\
p_2 = \dfrac{\alpha_{11}+\alpha_{01}}{ESS}, \\
\dfrac{\alpha_{11}\alpha_{00}-\alpha_{10}\alpha_{01}}{\sqrt{\alpha_{1+}\alpha_{+1}\alpha_{0+}\alpha_{+0}}} = \rho, \\
\alpha_{11}+\alpha_{10}+\alpha_{01}+\alpha_{00} = ESS,
\end{array}
\right.
\implies
\left\{
\begin{array}{lr}
\alpha_{11} = (\rho\sqrt{p_1p_2(1-p_1)(1-p_2)} + p_1p_2) \times ESS, \\
\alpha_{10} = p_1 \times ESS - \alpha_{11}, \\
\alpha_{01} = p_2 \times ESS - \alpha_{11}, \\
\alpha_{00} = ESS - (\alpha_{11}+\alpha_{10}+\alpha_{01}).
\end{array}
\right.
\end{equation}

\subsection{Calculating the Exceedance Probability}

Within each cohort, the stopping decision follows the same principles as single-arm monitoring (Section \ref{s:onearm}). Cohort $i$ stops for excess toxicity if $P(\theta_i > \theta_{0i} \mid n_1,n_2,k_1,k_2) > \tau$, where $\tau$ is a pre-specified cutoff. Based on Theorem \ref{thmm:thm2}, the exceedance probability for cohort 1 is
\begin{equation}
\begin{split}
P(\theta_1 > \theta_{01} \mid n_1,n_2,k_1,k_2) = & \int_{\theta_{01}}^1 [\theta_1 \mid n_1,n_2,k_1,k_2] d\theta_1 \\
= & \sum_{y=0}^{n_2} g_1(y) \bar{F}_{Be}(\theta_{01}; \alpha_{1+}+k_1+n_2-y,\: \alpha_{0+}+n_1-k_1+y),
\end{split}
\label{eq:ex_prop}
\end{equation}
where $\bar{F}_{Be}(x; \alpha,\beta)$ denotes the survival function of the $Be(\alpha,\beta)$ distribution. The exceedance probability for cohort 2 is obtained similarly.

When this probability exceeds the threshold $\tau$, we stop cohort $i$ and record $n_i$ and $k_i$. In our framework, we continue monitoring the other cohort, updating the calculation with the recorded $n_i$ and $k_i$. For example, if the probability first exceeds $\tau$ in cohort 1, we stop cohort 1 and record $n_1$ and $k_1$. We then continue monitoring cohort 2 with updated $n_2$ and $k_2$ and the known $n_1$ and $k_1$ until it stops or the maximum sample size $N$ is reached.

\subsection{Posteriors for Independent and Pooled Monitoring}

For independent monitoring, we essentially apply the single-arm Bayesian monitoring approach from Section \ref{s:onearm} to each cohort separately. Treating the two cohorts as distinct trials, and to maintain consistency with the correlated prior, we use the priors $[\theta_1] \sim \bm{Be}(\alpha_{11}+\alpha_{10}, \alpha_{01}+\alpha_{00})$ and $[\theta_2] \sim \bm{Be}(\alpha_{11}+\alpha_{01}, \alpha_{10}+\alpha_{00})$. Thus,
\begin{equation}
\begin{split}
[\theta_1 \mid n_1,k_1] \sim & \bm{Be}(\alpha_{11}+\alpha_{10}+k_1,\: \alpha_{01}+\alpha_{00}+n_1-k_1), \\
[\theta_2 \mid n_2,k_2] \sim & \bm{Be}(\alpha_{11}+\alpha_{01}+k_2,\: \alpha_{10}+\alpha_{00}+n_2-k_2).
\end{split}
\end{equation}

For pooled monitoring, we ignore the distinction between cohorts and apply single-arm Bayesian monitoring with a single toxicity probability for both cohorts. To maintain consistency with the other monitoring rules, we average the two priors, yielding $[\theta] \sim \bm{Be}((2\alpha_{11}+\alpha_{10}+\alpha_{01})/2,\: (2\alpha_{00}+\alpha_{01}+\alpha_{10})/2)$. Thus,
\begin{equation}
[\theta \mid n_1+n_2,k_1+k_2] \sim \bm{Be}\left(\frac{2\alpha_{11}+\alpha_{10}+\alpha_{01}}{2}+k_1+k_2,\: \frac{2\alpha_{00}+\alpha_{01}+\alpha_{10}}{2}+n_1+n_2-(k_1+k_2)\right).
\end{equation}

Using these posterior distributions, we can recalculate the posterior probabilities and thus the exceedance probabilities for independent and pooled monitoring using equation (\ref{eq:ex_prop}).

\subsection{Stopping Tables}
\label{s:stop_b}

The stopping boundary is determined using the exceedance probability (\ref{eq:ex_prop}). However, displaying the dynamic relationship between $k_1$ and $k_2$ in a limited space can be challenging. We therefore assume equal numbers of subjects in both cohorts at each assessment point. Stopping tables for the three approaches are provided in Table \ref{tab:stop}, with $\theta_{01}=\theta_{02}=0.2$, $ESS=3$, $\rho=0.5$, and $\tau=0.98$.

In Table \ref{tab:stop}, each cell indicates the number of toxicity events in cohort 1 ($k_1$) required to stop cohort 1, given the number of toxicity events in cohort 2 ($k_2$), assuming $n_1=n_2$ subjects are enrolled in each cohort. For example, when no toxicity events occur in cohort 2 and six patients are enrolled in each cohort, at least five toxicity events are needed to stop cohort 1 under both independent and correlated monitoring rules. Under pooled monitoring, six toxicity events would be required.

\begin{table}[htbp]
  \centering
  \caption{Stopping boundary table for toxicity ($\theta_{01}=\theta_{02}=0.2,\tau=0.98$)}
    \begin{tabular}{ccccccccccccc}
    \toprule
    \multicolumn{2}{p{3.645em}}{Stopping} & \multirow{2}{3cm}{$p_1=p_2$} & \multicolumn{10}{c}{$n_1=n_2=$} \\
    \multicolumn{2}{p{3.645em}}{Boundary} & & 1 & 2 & 3 & 4 & 5 & 6 & 7 & 8 & 9 & 10 \\
    \midrule
    \multirow{33}{2cm}{$k_2=$} & \multirow{3}{2cm}{0} & \multicolumn{1}{l}{0 (Independent)} & . & . & 3 & 4 & 4 & 5 & 5 & 5 & 6 & 6 \\
    & & \multicolumn{1}{l}{0.5 (Correlated)} & . & . & . & 4 & 4 & 5 & 5 & 6 & 6 & 6 \\
    & & \multicolumn{1}{l}{0.99 (Pooled)} & . & . & . & . & . & 6 & 7 & 7 & 8 & 8 \\
    \cmidrule{2-13}
    & \multirow{3}{2cm}{1} & \multicolumn{1}{l}{0 (Independent)} & . & . & 3 & 4 & 4 & 5 & 5 & 5 & 6 & 6 \\
    & & \multicolumn{1}{l}{0.5 (Correlated)} & . & . & 3 & 4 & 4 & 5 & 5 & 6 & 6 & 6 \\
    & & \multicolumn{1}{l}{0.99 (Pooled)} & . & . & . & 4 & 5 & 5 & 6 & 7 & 7 & 8 \\
    \cmidrule{2-13}
    & \multirow{3}{2cm}{2} & \multicolumn{1}{l}{0 (Independent)} & . & . & 3 & 4 & 4 & 5 & 5 & 5 & 6 & 6 \\
    & & \multicolumn{1}{l}{0.5 (Correlated)} & . & 2 & 3 & 4 & 4 & 4 & 5 & 5 & 6 & 6 \\
    & & \multicolumn{1}{l}{0.99 (Pooled)} & . & 2 & 3 & 3 & 4 & 5 & 5 & 6 & 6 & 7 \\
    \cmidrule{2-13}
    & \multirow{3}{2cm}{3} & \multicolumn{1}{l}{0 (Independent)} & . & . & 3 & 4 & 4 & 5 & 5 & 5 & 6 & 6 \\
    & & \multicolumn{1}{l}{0.5 (Correlated)} & . & . & 3 & 3 & 4 & 4 & 5 & 5 & 5 & 6 \\
    & & \multicolumn{1}{l}{0.99 (Pooled)} & . & . & 2 & 2 & 3 & 4 & 4 & 5 & 5 & 6 \\
    \cmidrule{2-13}
    & \multirow{3}{2cm}{4} & \multicolumn{1}{l}{0 (Independent)} & . & . & . & 4 & 4 & 5 & 5 & 5 & 6 & 6 \\
    & & \multicolumn{1}{l}{0.5 (Correlated)} & . & . & . & 3 & 3 & 4 & 4 & 5 & 5 & 5 \\
    & & \multicolumn{1}{l}{0.99 (Pooled)} & . & . & . & 2 & 2 & 3 & 3 & 4 & 4 & 5 \\
    \cmidrule{2-13}
    & \multirow{3}{2cm}{5} & \multicolumn{1}{l}{0 (Independent)} & . & . & . & . & 4 & 4 & 5 & 5 & 6 & 6 \\
    & & \multicolumn{1}{l}{0.5 (Correlated)} & . & . & . & . & 3 & 4 & 4 & 4 & 5 & 5 \\
    & & \multicolumn{1}{l}{0.99 (Pooled)} & . & . & . & . & 2 & 2 & 2 & 3 & 3 & 4 \\
    \cmidrule{2-13}
    & \multirow{3}{2cm}{6} & \multicolumn{1}{l}{0 (Independent)} & . & . & . & . & . & 4 & 5 & 5 & 6 & 6 \\
    & & \multicolumn{1}{l}{0.5 (Correlated)} & . & . & . & . & . & 4 & 4 & 4 & 5 & 5 \\
    & & \multicolumn{1}{l}{0.99 (Pooled)} & . & . & . & . & . & 2 & 2 & 2 & 3 & 3 \\
    \cmidrule{2-13}
    & \multirow{3}{2cm}{7} & \multicolumn{1}{l}{0 (Independent)} & . & . & . & . & . & . & 5 & 5 & 6 & 6 \\
    & & \multicolumn{1}{l}{0.5 (Correlated)} & . & . & . & . & . & . & 4 & 4 & 5 & 5 \\
    & & \multicolumn{1}{l}{0.99 (Pooled)} & . & . & . & . & . & . & 2 & 2 & 3 & 3 \\
    \cmidrule{2-13}
    & \multirow{3}{2cm}{8} & \multicolumn{1}{l}{0 (Independent)} & . & . & . & . & . & . & . & 5 & 6 & 6 \\
    & & \multicolumn{1}{l}{0.5 (Correlated)} & . & . & . & . & . & . & . & 4 & 5 & 5 \\
    & & \multicolumn{1}{l}{0.99 (Pooled)} & . & . & . & . & . & . & . & 3 & 3 & 3 \\
    \cmidrule{2-13}
    & \multirow{3}{2cm}{9} & \multicolumn{1}{l}{0 (Independent)} & . & . & . & . & . & . & . & . & 5 & 6 \\
    & & \multicolumn{1}{l}{0.5 (Correlated)} & . & . & . & . & . & . & . & . & 5 & 5 \\
    & & \multicolumn{1}{l}{0.99 (Pooled)} & . & . & . & . & . & . & . & . & 3 & 3 \\
    \cmidrule{2-13}
    & \multirow{3}{2cm}{10} & \multicolumn{1}{l}{0 (Independent)} & . & . & . & . & . & . & . & . & . & 6 \\
    & & \multicolumn{1}{l}{0.5 (Correlated)} & . & . & . & . & . & . & . & . & . & 5 \\
    & & \multicolumn{1}{l}{0.99 (Pooled)} & . & . & . & . & . & . & . & . & . & 4 \\
    \midrule
    \multicolumn{13}{r}{Note: $\theta_{01}=0.2, \theta_{02}=0.2, \tau=0.98$.} \\
    \bottomrule
    \end{tabular}
  \label{tab:stop}%
\end{table}%

\subsection{Frequentist Operating Characteristics of Stopping Rules}
\label{s:freq_op}

Investigators often require frequentist operating characteristics of the designed trial, including type I error, stopping probability, expected total number of enrolled patients, and expected number of events. We propose exact calculations using a recursive formula rather than simulation.

Since $n_1$ and $n_2$ are independent, so are $k_1$ and $k_2$, and $P(k_1,k_2 \mid n_1,n_2) = P(k_1 \mid n_1)P(k_2 \mid n_2)$. By the property of the binomial distribution,
\begin{equation}
P(k_i+1 \mid n_i+1) = \theta_i P(k_i \mid n_i) + (1-\theta_i) P(k_i+1 \mid n_i),
\label{eq:rec}
\end{equation}
where $0 \le k_i \le n_i$ and $P(0 \mid 0) = 1$.

Using the recursive formula (\ref{eq:rec}), we calculate $P(k_1,k_2 \mid n_1,n_2)$. Combined with the posterior probability of crossing the boundary for each $(k_1,k_2,n_1,n_2)$ as in equation (\ref{eq:post}), we obtain the frequentist operating characteristics for all three approaches.

\section{Comparison Among Three Monitoring Rules}

We compare the frequentist operating characteristics among the correlated, independent, and pooled stopping rules. We use a symmetric setting with $n_1=n_2=20$, $\theta_1=\theta_2=0.2$ in the prior, $ESS=3$, $\rho=0.5$, and $\tau=0.98$. This symmetric setup allows us to focus on the properties of one cohort, with the conclusions applying equally to the other. Without loss of generality, we present operating characteristics for cohort 1 when not specified. We demonstrate how to compute these quantities using the recursive algorithm in Section \ref{s:freq_op}.

\subsection{Type I Error}

Not surprisingly, the type I errors ($\alpha$) of the three approaches differ substantially when fixing $\tau=0.98$ and $N=20$. The type I error for cohort 1 is defined as
\begin{equation}
\alpha = P(\theta_1 \ge 0.2 \mid \theta_1=0.2, \theta_2).
\end{equation}

\begin{figure}[h]
\centerline{\includegraphics[height=8cm]{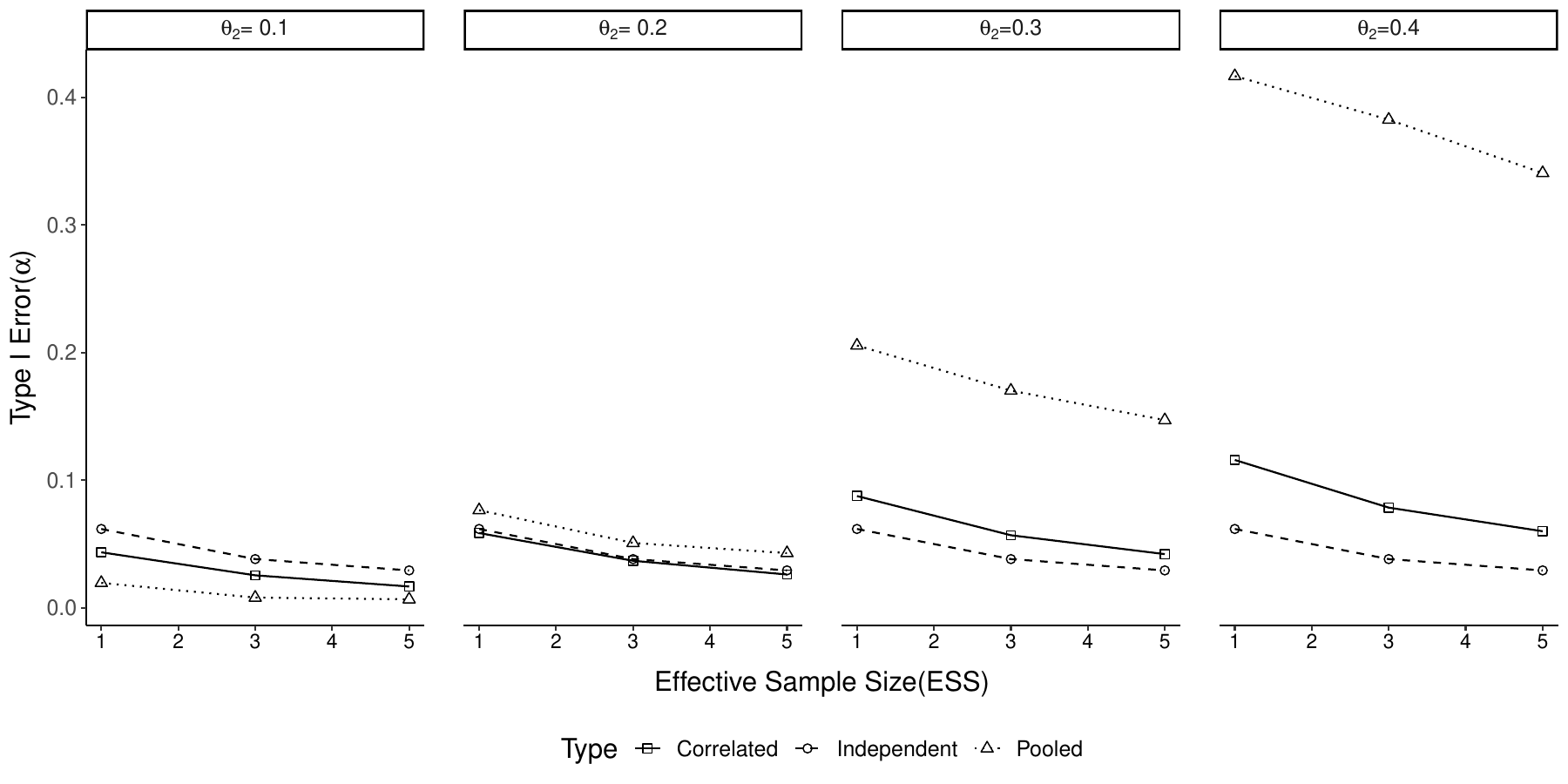}}
\caption{Type I error in cohort 1 ($\rho=0.5$, $N=20$, $\tau=0.98$)\label{fig:typeI}}
\end{figure}

Figure \ref{fig:typeI} shows the type I error panels separated by the actual toxicity in cohort 2 ($\theta_2$), with the effective sample size in the prior on the x-axis. Fixing $ESS$ and examining the relationship between $\alpha$ and $\theta_2$ across the three monitoring rules, we observe the following when $\theta_2=0.1$ (less than the toxicity threshold of 0.2): pooled monitoring yields the smallest (most conservative) type I error, independent monitoring performs worst but remains acceptable ($\le 0.1$), and correlated monitoring falls between them.

Our intuitive explanation is as follows. When pooled monitoring is applied, the average toxicity between the two cohorts affects the stopping decision. As $\theta_2$ increases, the pooled average moves farther from the upper bound $\theta_1=0.2$, making rejection of the null hypothesis more likely. Thus, $\alpha$ increases substantially as $\theta_2$ increases. In contrast, the posterior toxicity for cohort 1 under correlated monitoring is a weighted average of both cohorts, so the increasing trend is less pronounced. Independent monitoring is unaffected by the toxicity in cohort 2, so $\alpha$ remains constant across panels.

Turning to the effect of $ESS$, all lines in Figure \ref{fig:typeI} decrease as $ESS$ increases. This is expected because we set the prior means $\theta_1=\theta_2=0.2$, consistent with the null hypothesis. Consequently, more evidence is required to reject the null when the prior has greater weight.

\subsection{Stopping Probability}

To make all three rules comparable, we select $\tau$ to achieve a common type I error ($\alpha=0.1$). Controlling $\alpha=0.1$ yields the corresponding operating characteristics for all three approaches, which we present in the following sections.

\begin{figure}[h]
\centerline{\includegraphics[height=8cm]{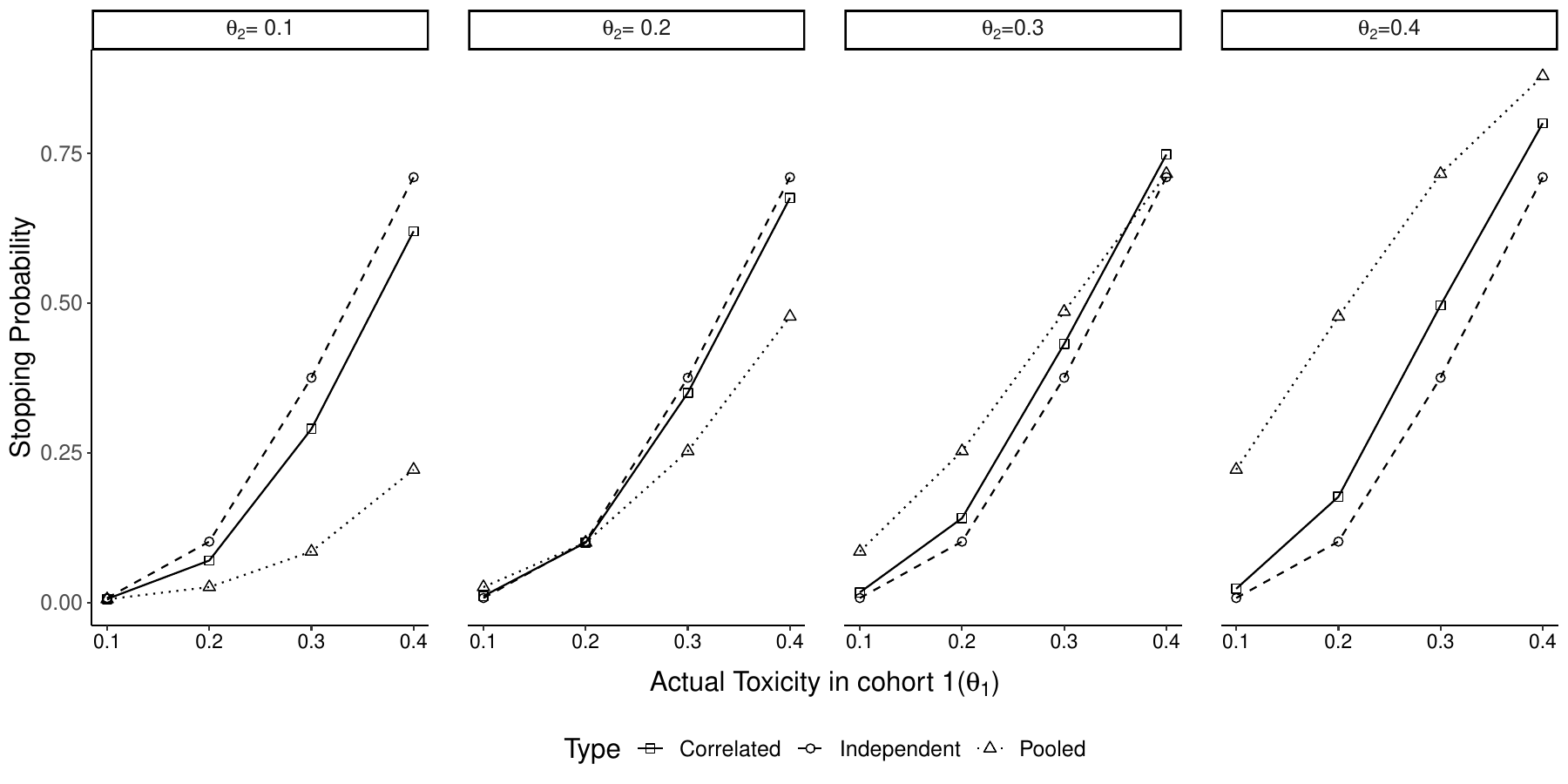}}
\caption{Probability of early stopping in cohort 1 ($\rho=0.5$, $N=20$, $\tau=0.98$, $ESS=3$)\label{fig:p_stop}}
\end{figure}

In Figure \ref{fig:p_stop}, the overall pattern of early stopping probability for cohort 1 increases with the toxicity probability across all three panels. Intuitively, more toxicity events are observed when true toxicity is high, making early stopping more likely.

First, comparing the three approaches across panels: the independent monitoring appears identical across panels, as expected, because the actual toxicity in cohort 2 ($\theta_2$) does not affect stopping in cohort 1 when monitored independently. The pooled curve shifts upward as $\theta_2$ increases. This occurs because higher toxicity in cohort 2 leads to more total events, even when cohort 1 toxicity remains constant. Since pooled monitoring treats both cohorts identically, cohort 1 stops more frequently. The correlated approach shifts upward, but less than the pooled method. Whether this upward shift is beneficial depends on the specific context.

Next, comparing within each panel: we expect the stopping probability for cohort 1 to be low when the actual toxicity ($\theta_1$) is 0.2 or less, and high when $\theta_1$ exceeds 0.2. Thus, independent monitoring performs best when $\theta_2=0.1$ or $0.2$. However, the best approach is less clear when $\theta_2=0.3$ or $0.4$. In the third panel ($\theta_2=0.2$), independent monitoring produces fewer stops than the other two approaches when $\theta_1=0.2$, but correlated monitoring stops more frequently than the others when $\theta_1=0.3$, with independent monitoring intermediate. A similar pattern appears in the fourth panel.

We expect independent monitoring to be preferable when toxicities differ substantially between cohorts, while pooled monitoring is more appropriate when toxicities are similar or identical. This expectation aligns with Figure \ref{fig:p_stop}, with one notable exception. In the second panel where $\theta_1=0.4$, correlated monitoring stops more frequently than independent monitoring even though $\theta_1$ and $\theta_2$ differ. This observation suggests that correlated monitoring may be advantageous in certain scenarios.

\subsection{Expected Total Number of Enrolled Patients}

\begin{figure}[h]
\centerline{\includegraphics[height=8cm]{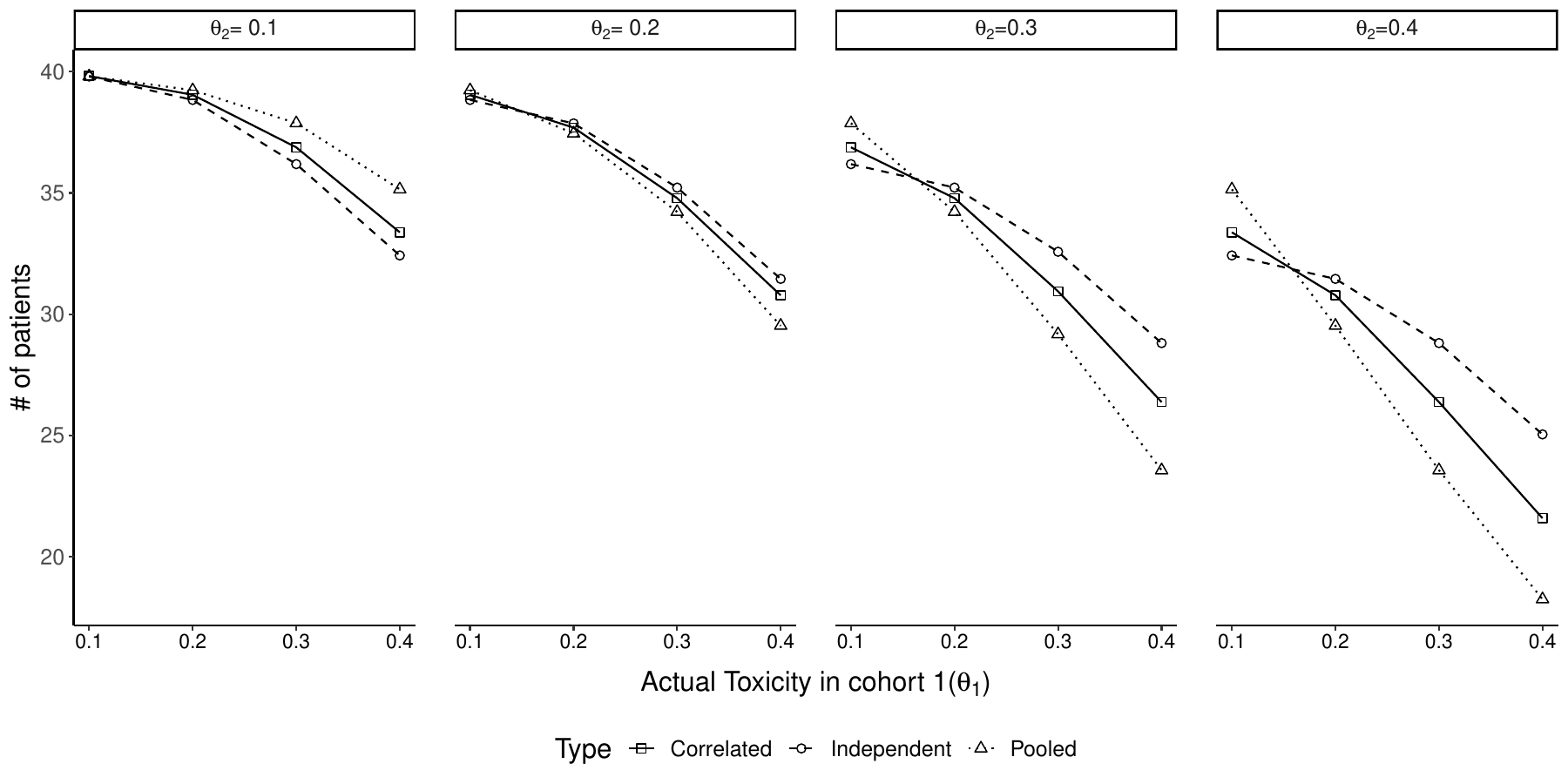}}
\caption{Expected number of patients enrolled at stopping ($\rho=0.5$, $N=20$, $\tau=0.98$, $ESS=3$)\label{fig:patients}}
\end{figure}

Considering patient safety and trial costs, fewer total enrolled patients is preferable. The overall pattern in Figure \ref{fig:patients} is clear: the total number of subjects required decreases as toxicity increases, because higher toxicity makes detection easier and early termination more likely. Independent monitoring enrolls the fewest patients when $\theta_2=0.1$, but this advantage reverses when $\theta_1, \theta_2 \ge 0.2$. Pooled monitoring performs better when $\theta_1, \theta_2 \ge 0.2$ and closely resembles correlated monitoring when $\theta_2=0.2$. Across most scenarios, correlated monitoring provides a balanced effect on sample size.

\subsection{Expected Number of Events}

From a safety perspective, we prefer fewer toxicity events regardless of true toxicity. We examine two related quantities: the total number of events when the trial ends (whether due to early stopping or reaching maximum sample size), and the number of events at early stopping in cohort 1.

\begin{figure}[h]
\centerline{\includegraphics[height=8cm]{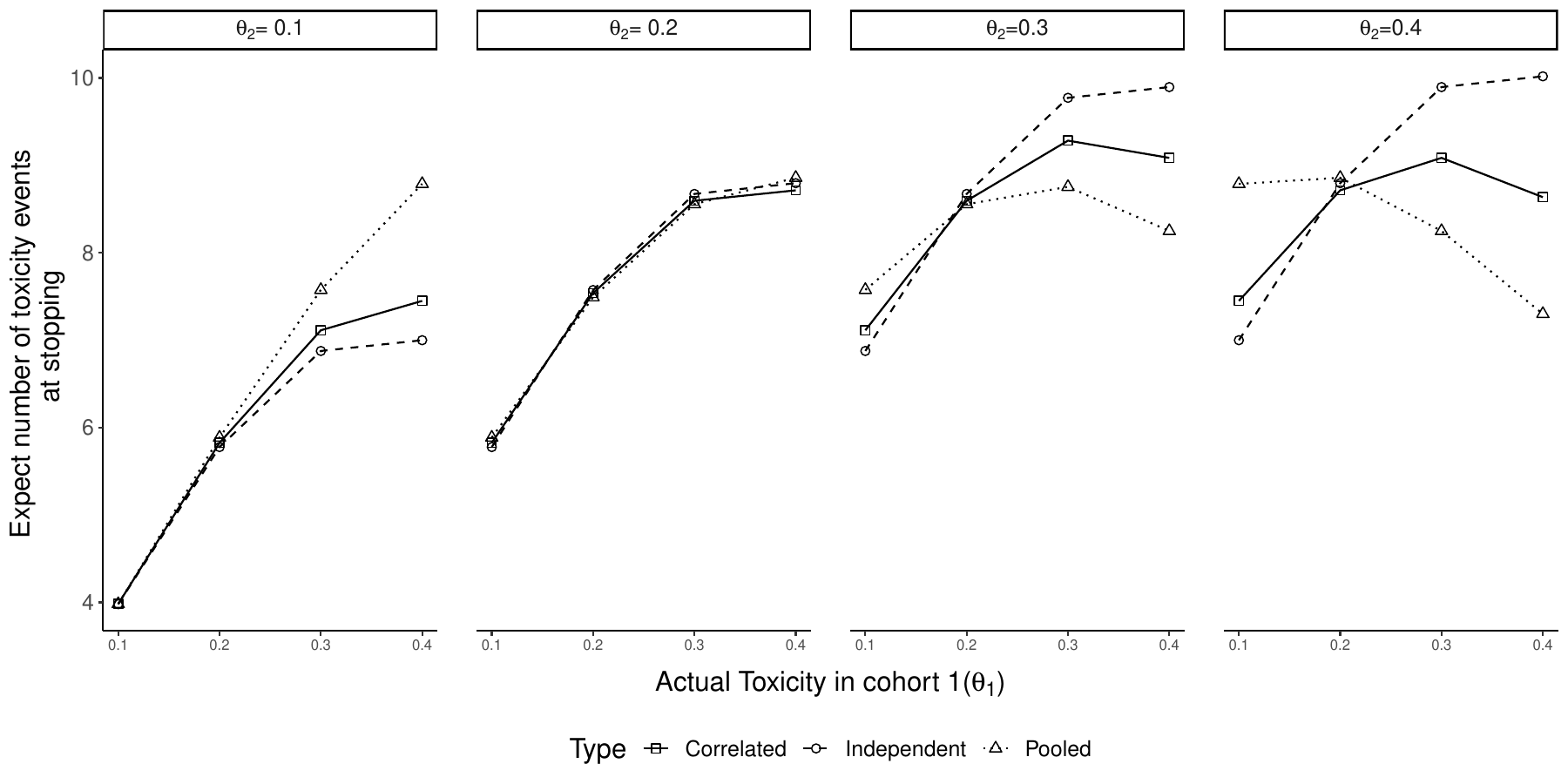}}
\caption{Expected number of events at stopping ($\rho=0.5$, $N=20$, $\tau=0.98$, $ESS=3$)\label{fig:events}}
\end{figure}

In Figure \ref{fig:events}, regardless of stopping rules, both cohorts should reach the maximum sample size $N_1=N_2=20$ when true toxicities are acceptable ($\theta_1=\theta_2=0.1$). Thus, the expected total number of events is $N_1 \theta_1 + N_2 \theta_2 = 4$. Extending this reasoning to different values of $\theta_1, \theta_2 \le 0.2$, when $\theta_1=0.2$ and $\theta_2=0.1$, the expected quantity is approximately $20 \times 0.2 + 20 \times 0.1 = 6$. The difference from 6 may arise because type I error allows some chance of early stopping even when true toxicity is acceptable. Consequently, the expected number of toxicity events is slightly less than 6. The general trend shows increasing expected events as toxicity probabilities increase, with some exceptions for correlated and pooled monitoring when $\theta_1=\theta_2=0.4$. When true toxicities are extremely high, the trial may end very early, and the reduced enrollment mitigates the effect of high toxicity on event counts. Notably, all three rules perform similarly in the second panel ($\theta_2=0.2$). Correlated monitoring consistently provides intermediate effects across all panels.

\begin{figure}[h]
\centerline{\includegraphics[height=8cm]{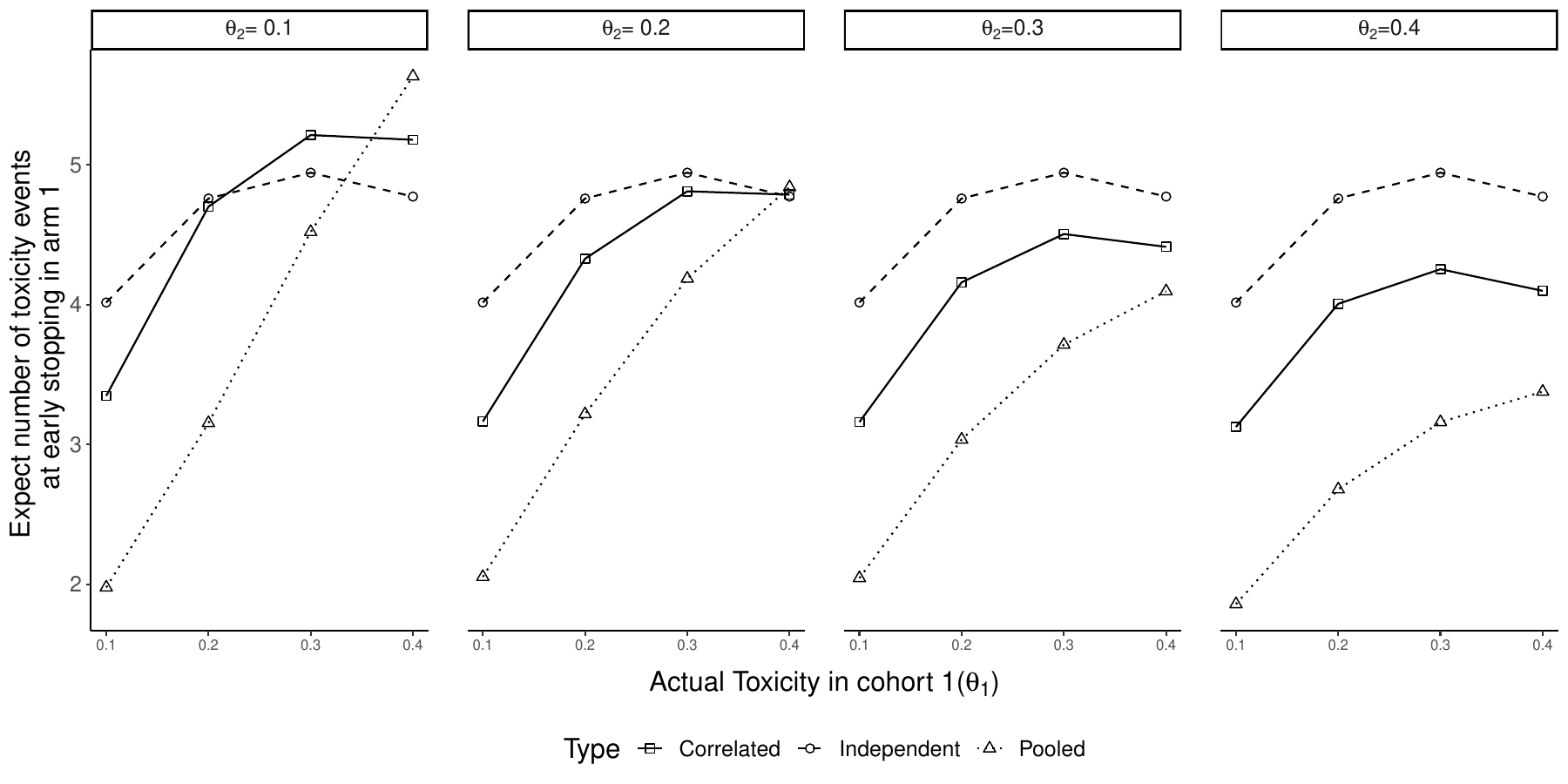}}
\caption{Expected number of events at early stopping in cohort 1 ($\rho=0.5$, $N=20$, $\tau=0.98$, $ESS=3$)\label{fig:early.events.1}}
\end{figure}

We also present the expected number of toxicity events at early stopping in cohort 1 for the three approaches. These values are collected only when cohort 1 stops early and exclude cases where the trial reaches the sample size limit. The dashed reference lines in Figure \ref{fig:early.events.1} are identical across panels because this quantity depends only on cohort 1, independent of cohort 2. These values reflect the difficulty of stopping cohort 1.

\section{Discussion}
\label{s:discuss}

As discussed earlier, two-cohort toxicity monitoring presents challenges with current methods. Independent monitoring ignores the similarity between cohorts, while pooled monitoring neglects heterogeneity between them. Both methods have scenarios where they outperform our joint monitoring rule, as demonstrated in the previous section. However, the true relationship between toxicities in two cohorts is often uncertain in practice, making it difficult to choose the most appropriate monitoring rule a priori for optimal operating characteristics.

Our method provides an alternative approach for toxicity monitoring in two-cohort trials. By introducing the bivariate beta prior into the Bayesian framework, we achieve a compromise between the two existing rules: neither treating the cohorts as independent nor identical. The degree of correlation can be adjusted through the effective sample size ($ESS$) and correlation coefficient ($\rho$). Higher prior correlation leads to greater posterior correlation between cohorts, with the same applying to ESS. Additionally, compared to traditional simulation methods, we obtain exact operating characteristics rather than simulated estimates in a relatively efficient manner. This approach yields more stable properties across different settings. Although it is not optimal in all scenarios, it consistently performs near the best option.

Several limitations should be noted. Unlike the deterministic boundaries provided by other approaches, our method produces a more dynamic stopping boundary that requires collaboration with a statistician to monitor toxicity throughout the trial. This may pose challenges for safety monitoring boards. To address this limitation, we have developed an R package to facilitate implementation.

Several extensions are possible, including allowing more than two cohorts, using a distribution on the prior correlation instead of a single pre-specified value $\rho$, and evaluating different toxicity thresholds.

\bibliographystyle{biom}
\bibliography{references}

\pagebreak
\begin{appendices}
\subsection*{Proof of Theorem 1}

Starting from equation (\ref{eq:post}), we have
\newline
\scalebox{0.8}{
\begin{minipage}{\textwidth}
\begin{align*}
K(\theta_1,\theta_2) = & \int_\Omega u^{\alpha_{11}-1}(\theta_1-u)^{\alpha_{10}-1}(\theta_2-u)^{\alpha_{01}-1}(1-\theta_1-\theta_2+u)^{\alpha_{00}-1} \theta_1^{k_1}(1-\theta_1)^{n_1-k_1} \theta_2^{k_2}(1-\theta_2)^{n_2-k_2} du \\
= & \int_\Omega u^{\alpha_{11}-1}(\theta_1-u)^{\alpha_{10}-1}(\theta_2-u)^{\alpha_{01}-1}(1-\theta_1-\theta_2+u)^{\alpha_{00}-1} (u+\theta_1-u)^{k_1}(\theta_2-u + 1-\theta_1-\theta_2+u)^{n_1-k_1} \\
& \cdot (u+\theta_2-u)^{k_2}(\theta_1-u + 1-\theta_2-\theta_1+u)^{n_2-k_2} du \\
= & \int_\Omega u^{\alpha_{11}-1}(\theta_1-u)^{\alpha_{10}-1}(\theta_2-u)^{\alpha_{01}-1}(1-\theta_1-\theta_2+u)^{\alpha_{00}-1} \left\{\sum_{x_1=0}^{k_1} {k_1 \choose x_1} u^{x_1} (\theta_1-u)^{k_1-x_1}\right\} \\
& \cdot \left\{\sum_{x_2=0}^{n_1-k_1} {{n_1-k_1} \choose x_2} (\theta_2-u)^{x_2} (1-\theta_1-\theta_2+u)^{n_1-k_1-x_2}\right\} \left\{\sum_{y_1=0}^{k_2} {k_2 \choose y_1} u^{y_1} (\theta_2-u)^{k_2-y_1}\right\} \\
& \cdot \left\{\sum_{y_2=0}^{n_2-k_2} {{n_2-k_2} \choose y_2} (\theta_1-u)^{y_2} (1-\theta_1-\theta_2+u)^{n_2-k_2-y_2}\right\} du \\
= & \sum_{x_1=0}^{k_1} \sum_{x_2=0}^{n_1-k_1} \sum_{y_1=0}^{k_2} \sum_{y_2=0}^{n_2-k_2} \left\{ {k_1 \choose x_1} {{n_1-k_1} \choose x_2} {k_2 \choose y_1} {{n_2-k_2} \choose y_2} \right. \\
& \left. \cdot \int_\Omega u^{\alpha_{11}-1+x_1+y_1} (\theta_1-u)^{\alpha_{10}-1+k_1-x_1+y_2} (\theta_2-u)^{\alpha_{01}-1+x_2+k_2-y_1} (1-\theta_1-\theta_2+u)^{\alpha_{00}-1+n_1-k_1-x_2+n_2-k_2-y_2} du \right\} \\
= & \sum_{x_1=0}^{k_1} \sum_{x_2=0}^{n_1-k_1} \sum_{y_1=0}^{k_2} \sum_{y_2=0}^{n_2-k_2} {k_1 \choose x_1} {{n_1-k_1} \choose x_2} {k_2 \choose y_1} {{n_2-k_2} \choose y_2} \bm B(\bm \alpha') \\
& \cdot f_{{\bm Be}_2}(\alpha_{11}+x_1+y_1,\: \alpha_{10}+k_1-x_1+y_2,\alpha_{01}+x_2+k_2-y_1,\:\alpha_{00}+n_1-k_1-x_2+n_2-k_2-y_2).
\end{align*}
\end{minipage}
}

\subsection*{Proof of Theorem 2}

\begin{equation}
\begin{split}
g_1(y) = & c_1 \Gamma(\alpha_{1+}+k_1+n_2-y) \Gamma(\alpha_{0+}+n_1-k_1+y) \sum_{y_2=\max(0, y-k_2)}^{\min(n_2-k_2, y)} {k_2 \choose {y-y_2}} {{n_2-k_2} \choose y_2} \\
& \times B(\alpha_{00}+y_2, \alpha_{01}+y-y_2) B(\alpha_{10}+n_2-k_2-y_2, \alpha_{11}+k_2-y+y_2).
\end{split}
\end{equation}

\begin{equation}
\begin{split}
g_2(x) = & c_2 \Gamma(\alpha_{+1}+k_2+n_1-x) \Gamma(\alpha_{+0}+n_2-k_2+x) \sum_{x_2=\max(0, x-k_1)}^{\min(n_1-k_1, x)} {k_1 \choose {x-x_2}} {{n_1-k_1} \choose x_2} \\
& \times B(\alpha_{00}+x_2, \alpha_{10}+x-x_2) B(\alpha_{01}+n_1-k_1-x_2, \alpha_{11}+k_1-x+x_2).
\end{split}
\end{equation}

Here, $c_1$ and $c_2$ are scaling constants.

Since $x_1, x_2, y_1, y_2$, and $\theta_1$ are constant with respect to $\theta_2$, and the sum of integrals equals the integral of the sum, we can move the integral inside the summation when integrating the joint posterior (\ref{eq:joint}) over $\theta_2$. The marginal distribution of $\theta_1$ is $\bm{Be}(\alpha_{1+}, \alpha_{0+})$ by the property of $\bm{Be}_2(\alpha_{11}, \alpha_{10}, \alpha_{01}, \alpha_{00})$. Applying the same argument with $\bm\alpha$ replaced by $\bm\alpha'$ completes the proof.

This completes the proof. A similar derivation holds for $\theta_2$.

Moreover,
\begin{equation}
\begin{aligned}
B(\bm \alpha') = & \frac{\prod_{i=1}^4 \Gamma(\alpha_i')}{\Gamma(\sum_{i=1}^4 \alpha_i')} \\
= & \frac{\Gamma(\alpha_{11}+k_1-x_1+k_2-y_1) \Gamma(\alpha_{10}+x_1+n_2-k_2-y_2) \Gamma(\alpha_{01}+n_1-k_1-x_2+y_1) \Gamma(\alpha_{00}+x_2+y_2)}{\Gamma(\alpha_{11}+\alpha_{10}+\alpha_{01}+\alpha_{00}+n_1+n_2)} \\
\propto & \Gamma(\alpha_{11}+k_1-x_1+k_2-y_1) \Gamma(\alpha_{10}+x_1+n_2-k_2-y_2) \Gamma(\alpha_{01}+n_1-k_1-x_2+y_1) \Gamma(\alpha_{00}+x_2+y_2).
\end{aligned}
\end{equation}

The term $\Gamma(\alpha_{11}+k_1-x_1+k_2-y_1) \Gamma(\alpha_{10}+x_1+n_2-k_2-y_2)$ is the kernel of a beta-binomial distribution with $n=k_1$, $k=x_1$, $\alpha=n_2-k_2-y_2+\alpha_{10}$, and $\beta=k_2-y_1+\alpha_{11}$. Similarly, $\Gamma(\alpha_{01}+n_1-k_1-x_2+y_1) \Gamma(\alpha_{00}+x_2+y_2)$ is the kernel with $n=n_1-k_1$, $k=x_2$, $\alpha=\alpha_{00}+y_2$, and $\beta=\alpha_{01}+y_1$.

Note that if $k \sim \text{Beta-Binomial}(n,\alpha,\beta)$, then
\begin{align*}
d_{\text{betabinom}}(k) = & {n \choose k} \frac{B(k+\alpha, n-k+\beta)}{B(\alpha,\beta)} \\
= & {n \choose k} \frac{\frac{\Gamma(k+\alpha)\Gamma(n-k+\beta)}{\Gamma(n+\alpha+\beta)}}{\frac{\Gamma(\alpha)\Gamma(\beta)}{\Gamma(\alpha+\beta)}} \\
= & \frac{\Gamma(\alpha+\beta)}{\Gamma(n+\alpha+\beta)\Gamma(\alpha)\Gamma(\beta)} {n \choose k} \Gamma(k+\alpha) \Gamma(n-k+\beta).
\end{align*}

Thus,
\begin{align*}
{n \choose k} \Gamma(k+\alpha) \Gamma(n-k+\beta) = d_{\text{betabinom}}(k) \frac{\Gamma(n+\alpha+\beta)\Gamma(\alpha)\Gamma(\beta)}{\Gamma(\alpha+\beta)} = d_{\text{betabinom}}(k) \Gamma(n+\alpha+\beta) B(\alpha,\beta).
\end{align*}

The full derivation proceeds as follows:
\begin{align*}
p(\theta_1 \mid n_1,n_2,k_1,k_2) \propto & \sum_{x_1=0}^{k_1} \sum_{x_2=0}^{n_1-k_1} \sum_{y_1=0}^{k_2} \sum_{y_2=0}^{n_2-k_2} \Big\{ {k_1 \choose x_1} {{n_1-k_1} \choose x_2} {k_2 \choose y_1} {{n_2-k_2} \choose y_2} \\
& \quad \cdot \Gamma(\alpha_{11}+k_1-x_1+k_2-y_1) \Gamma(\alpha_{10}+x_1+n_2-k_2-y_2) \\
& \quad \cdot \Gamma(\alpha_{01}+n_1-k_1-x_2+y_1) \Gamma(\alpha_{00}+x_2+y_2) \\
& \quad \cdot d_{\text{beta}}(\theta_1,\alpha_{11}+k_1-y_1+\alpha_{10}+n_2-y_2,\: \alpha_{01}+n_1-k_1+y_1+\alpha_{00}+y_2) \Big\} \\
\propto & \sum_{y_1=0}^{k_2} \sum_{y_2=0}^{n_2-k_2} \Big\{ \Gamma(k_1+n_2-y_2+\alpha_{10}-y_1+\alpha_{11}) B(n_2-k_2-y_2+\alpha_{10}, k_2-y_1+\alpha_{11}) \\
& \quad \cdot \Gamma(n_1-k_1+\alpha_{00}+y_2+\alpha_{01}+y_1) B(\alpha_{00}+y_2, \alpha_{01}+y_1) {k_2 \choose y_1} {{n_2-k_2} \choose y_2} \\
& \quad \cdot d_{\text{beta}}(\theta_1,\alpha_{11}+k_1-y_1+\alpha_{10}+n_2-y_2,\: \alpha_{01}+n_1-k_1+y_1+\alpha_{00}+y_2) \\
& \quad \cdot \sum_{x_1=0}^{k_1} d_{\text{betabinom}}(x_1 \mid n=k_1, k=x_1, \alpha=n_2-k_2-y_2+\alpha_{10}, \beta=k_2-y_1+\alpha_{11}) \\
& \quad \cdot \sum_{x_2=0}^{n_1-k_1} d_{\text{betabinom}}(x_2 \mid n=n_1-k_1, k=x_2, \alpha=\alpha_{00}+y_2, \beta=\alpha_{01}+y_1) \Big\} \\
\propto & \sum_{y=0}^{n_2} \Big\{ \Gamma(k_1+n_2-y+\alpha_{10}+\alpha_{11}) \Gamma(n_1-k_1+\alpha_{00}+y+\alpha_{01}) \\
& \quad \cdot d_{\text{beta}}(\theta_1,\alpha_{11}+k_1-y+\alpha_{10}+n_2,\: \alpha_{01}+n_1-k_1+y+\alpha_{00}) \\
& \quad \cdot \sum_{y_2=\max(0, y-k_2)}^{\min(n_2-k_2, y)} B(n_2-k_2-y_2+\alpha_{10}, k_2-y+y_2+\alpha_{11}) \\
& \quad \cdot B(\alpha_{00}+y_2, \alpha_{01}+y-y_2) {k_2 \choose {y-y_2}} {{n_2-k_2} \choose y_2} \Big\}.
\end{align*}

This completes the proof. A similar derivation holds for $\theta_2$.
\end{appendices}
\end{document}